\documentclass[12pt]{iopart}
\usepackage{graphicx}

\begin{document}
\title[Theoretical study of Frenkel pairs recombination in SiC]{Theoretical study of the recombination of Frenkel pairs in irradiated silicon carbide}

\author{Guillaume Lucas and Laurent Pizzagalli}
\ead{Laurent.Pizzagalli@univ-poitiers.fr} 
\address{Laboratoire de M{\'e}tallurgie Physique, CNRS UMR 6630, Universit{\'e}
de Poitiers,  B.P~30179, F-86962~Futuroscope Chasseneuil Cedex, France}

\pacs{61.72.Cc,61.72.Ji,61.80.Jh}
\submitto{\JPCM}

\begin{abstract}
The recombination of Frenkel pairs resulting from low energy recoils in 3C-SiC has been investigated using first principles and Nudged Elastic Band calculations. 
Several recombination mechanisms have been obtained, involving direct interstitial migration, atoms exchange, or concerted displacements, with activation energies ranging from 0.65~eV to 1.84~eV. These results are in agreement with experimental activation energies. We have determined the lifetime of the $V_{Si}+Si_{TC}$ Frenkel pair, by computing phonons and the Arrhenius prefactor. The vibrational contributions to the free energy barrier have been shown to be negligible in that case. 
\end{abstract}

\section{Introduction}

Silicon carbide has been extensively studied and used in various applications, due to its unique physical, chemical and mechanical properties~\cite{Cho97MRSB}. In electronics, SiC is a possible replacement for silicon in high temperature, high power and high frequency devices. In radioactive environment, possible uses concern fusion reactors, confinement matrices or spatial electronics. A good knowledge of the crystalline SiC behaviour during and after irradiation is a prerequisite for all these applications. In fact, during irradiation, lattice atoms are displaced, resulting in the formation of structural defects such as interstitials and vacancies. The damage is accumulating in the material, possibly leading to mechanical and electrical properties deterioration, and even amorphization in the case of large doses or low temperatures \cite{Wen98NIMB}. Also, only ion implantation is a viable option to dope SiC-based electronic devices, since dopants have high migration energies in silicon carbide, preventing the use of conventional thermal diffusion techniques. The accumulation of damage in SiC due to ion implantation has been largely studied \cite{Sko96NIMPR,Hal02NIMPR}. 

In order to anneal these defects and recover a good crystalline quality, thermal treatments have been applied during or after irradiation. The defects concentration will change according to temperature and irradiation parameters, as a dynamic process of defect creation and recombination. To better understand the materials behaviour or the crystal recovery, after irradiation or annealing treatement, knowledge of the stability and mobility properties of defects is required. In particular, important quantities are the activation energies for recombining vacancies and interstitials from the Frenkel pairs formed in irradiated materials. Experimentally, it has been shown that distinct recovery stages exist as a function of temperature and of the level of damage, for 4H-SiC \cite{Zha02JAP}. In the case of 6H-SiC, isochronal and isothermal annealings indicate three recovery stages as a function of temperature, with activation energies $0.3\pm0.15$~eV (150-300~K), $1.3\pm0.25$~eV (450-550~K), and $1.5\pm0.3$~eV (570-720~K) \cite{Web00NIMB,Web01NIMB}. This last result is in accordance with a study of dynamic annealing in 4H-SiC, yielding an activation energy of $1.3$~eV for recovery with a temperature ranging from 350~K to 430~K \cite{Kuz03JAP}. Finally, in the high temperature domain, Itoh \textit{et al.} have shown that irradiation-induced centers in 3C-SiC are fully annealed above 1020~K, with a measured activation energy of $2.2\pm0.3$~eV~\cite{Ito89JAP}.

The available informations are not complete enough to draw a full and comprehensible picture of the defects annealing and crystal recovery of silicon carbide. It is worth to note that cascade simulations reveal that most of the formed defects due to irradiation are Frenkel pairs with very few clustered defects and antisites \cite{Per00JNM, Dev01JAP}. Two mechanims are then mainly responsible for defect annealings, the first being short range recombination of Frenkel pairs, and the second long range migration of point defects. The latter has been largely studied, and the activation energies for single point defects have been calculated using quantum mechanical approaches~\cite{Boc03PRB,RauPHD,Sal04JNM}. Less knowledge has been accumulated for the former process, principally because of the large number of possible Frenkel pair configurations. Using classical molecular dynamics, Gao and Weber have determined that activation energies for Frenkel pairs recombination range from 0.22~eV to 1.6~eV \cite{Gao03JAP}. Another study yields an energy barrier of 1.16~eV for the recombination of $V_{Si}$ and $SiSi$ dumbbell \cite{Mal00JNM,Mal02PRB}. These results suffer from the poor accuracy of classical potentials for describing transition states. Using more accurate Density Functional Theory (DFT) methods and transition states calculation techniques, Bockstedte \textit{et al.}~\cite{Boc04PRB} and Rauls~\cite{RauPHD} have investigated the recombination mechanisms and the associated activation energies. However, the few selected configurations have been built by associating an interstitial with a vacancy in an arbitrary way.

Using first principles DFT calculations, we have investigated the recombination of Frenkel pairs with short interstitial-vacancy distances in 3C-SiC. The recombination mechanism and the associated activation energies have been determined with the Nudged Elastic Band (NEB) method~\cite{Jon98WS,Hen00JCP}. The Frenkel pairs considered in this work have been obtained during ab initio molecular dynamics of low energy recoils~\cite{Luc05PRB}. These configurations are then supposed to provide a realistic description of Frenkel pairs occurring in irradiated SiC, in the case of a low density of defects. After a short report of our results on the stability of Frenkel pairs, the calculated mechanisms and energy barriers for recombination are described. The lifetime of one Frenkel pair has been estimated, taking into account vibrational entropy contributions. Finally we discuss our results in relation with experiments and previous studies. 

\section{Methods}

We employed the plane-wave pseudopotential code PWscf included in the Quantum-Espresso package~\cite{Quantum-Espresso}. Our calculations have been performed in the framework of Density Functional Theory~\cite{Hoh64PR,Koh65PR} and the Generalized Gradient Approximation (GGA-PBE) as parametrized by Perdew, Burke and Ernzerof~\cite{Per96PRL}. Vanderbilt ultrasoft pseudopotentials have been used for describing electron-ion interactions. We found that with these pseudopotentials, a plane wave basis with an energy cutoff of 25~Ry is enough to get difference energies converged to 0.01~eV. The silicon carbide lattice parameter is $a_0=4.382$~\AA. Defect configurations are modelled with periodically repeated supercells. Single defects and Frenkel pairs stability have been calculated in cells encompassing 216 and 64 atoms. The Brillouin zone sampling was performed using the $\Gamma$ point only for large cells, and a $4\times4\times4$ special k-points mesh for small cells \cite{Mon76PRB}. For recombination investigations, smaller cells including 64 or 96 atoms have been considered, as well as $4\times4\times4$ (64 atoms) or $2\times2\times2$ (96 atoms) special k-points meshs, in order to keep reasonable computational times. In this work, we considered defects in their neutral charge state. 

Two different approachs may be used for studying the recombination of Frenkel pairs. The first, based on molecular dynamics and Arrhenius plots, allows to determine recombination without prior knowledge of the path. Unfortunately, energy barriers for Frenkel pairs recombination in silicon carbide are expected to be large. The probability to obtain a successful event is then low, and very long molecular dynamics runs at high temperature are then required, preventing first principles molecular dynamics calculations. It is also not clear whether the harmonic approximation remains valid for high temperatures, or whether the right recombination mechanisms are activated. The second method, employed in this work, is the determination of the minimum energy path for recombination. We used the NEB technique~\cite{Jon98WS}, with three images between initial (Frenkel pair) and final (perfect crystal) states. The climbing image algorithm was also used to determine precisely the transition state~\cite{Hen00JCP}. While such a technique allows a precise calculation of the recombination mechanism as well as the associated activation energy, a guess of the initial path is required. Non-trivial mechanisms are then difficult to obtain. 
  
\section{Frenkel pairs stability}

\begin{table}
\begin{center}
\begin{tabular}[h]{cccccc}
\hline
   & & \multicolumn{2}{c}{216 atoms} & \multicolumn{2}{c}{64 atoms} \\
  & $d_{FP}$ & $E_f$ & $\Delta E$ & $E_f$ & $\Delta E$ \\ 
\hline
$V_C+CC_{[100]}$ & 0.85 & 9.90 & -0.03 & 10.69 & -0.17\\
$V_C+CSi_{[100]}$ & 0.5 & 6.69 & -3.24 & 7.06 & -3.80\\
$V_C+CSi_{[100]}$ & 0.95 & 9.96 & 0.03 & 10.86 & 0.00\\
$^*V_{Si}+Si_{TC}$ & 1.5 & 14.08 & -0.44 & 14.92 & -1.56\\
$V_{Si}+Si_{TC}$ & 0.9 & 13.46 & -1.06 & 13.26 & -3.22\\
\hline
\end{tabular}
\caption{Calculated formation energy $E_f$ (in eV) of several Frenkel pairs, for two cell sizes. $d_{FP}$ (in $a_0$) is the interstitial-vacancy distance, 
and $\Delta E$ is the formation energy difference between a Frenkel pair and individual point defects. A larger cell (96 atoms) has been considered for the configuration $^*$ due to the large vacancy-interstitial distance.} \label{FP}
\end{center}
\end{table}

The Figures~\ref{FP1}-\ref{FP5} shows the considered Frenkel pairs. Structural properties of these defects have already been described in a previous publication~\cite{Luc06NIMB}. In the Table~\ref{FP} the defect formation energies are reported, calculated according to the usual formalism~\cite{Zha91PRL}, for two different cell sizes. Among all Frenkel pairs, $V_C+CSi_{[100]}$ has the lowest formation energy. More generally, Frenkel pairs including C vacancies have the lowest formation energies, as expected regarding the low formation energies of C defects (both vacancy and interstitials) compared to Si defects~\cite{Luc06NIMB}. 
It would be interesting to compare these energies, obtained in the framework of GGA, with previous calculations performed using Local Density Approximation (LDA). Unfortunately, the Frenkel pair formation energies are often not reported in the available literature. Rauls points to energies of 10-11~eV for $V_C+CC$ pairs depending on the distance~\cite{RauPHD}, in agreement with our results. 

In the Table~\ref{FP} is also reported the difference between the formation energy of Frenkel pairs and the sum of the formation energies of individual defects. This quantity allows to compare the Frenkel pair stability against dissociation. Almost all differences are negative, suggesting that Frenkel pairs are more stable than single defects. Also, large energy differences are mostly associated with short vacancy-interstitial distances, indicating an attractive interaction between vacancies and interstitials~\cite{Luc06NIMB}. 

One important aspect is the influence of cell size on the stability of Frenkel pairs. In the Table~\ref{FP1} is reported the formation energies for cells including 216 or 64 atoms. Differences as large as 0.9~eV are obtained. We have observed a similar effect for single point defects, with larger formation energies for the carbon vacancy, and interstitials. The biggest difference is obtained for silicon interstitials, suggesting that a 64 atoms cell is too small to allow the full lattice distortion around the defect. Although present, this effect is less pronounced for Frenkel pairs, since there is a better accomodation of strains thanks to the vacancy. Using small cells, one may then expect an overestimation of the energy barrier for creating the Frenkel pair from the perfect crystal. Nevertheless, it is reasonable to assume much smaller differences for the recombination energy barrier, since similar lattice distortions are expected for the Frenkel pair configuration and the transition state. At most, lattice distortion should be larger for the transition state, leading to a slight overestimation of the recombination barriers.   

\section{Frenkel pairs recombination}

\begin{figure}
\begin{center}
\includegraphics*[width=8cm]{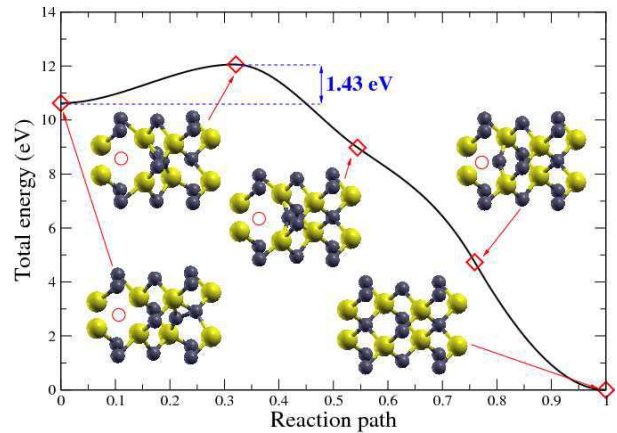}
\caption{Minimum energy path for the recombination of the Frenkel pair $V_C+CC_{[100]}$ ($d_{FP}=0.85 a_0$). The insets show the relaxed configurations along the minimum energy path. Light (dark) spheres are silicon (carbon) atoms, while the empty circle marks the position of the vacancy.} \label{FP1}
\end{center}
\end{figure}

The Figure~\ref{FP1} shows the recombination path and the energy barrier for the Frenkel pair $V_C+CC_{[100]}$ ($d_{FP}=0.85 a_0$). The carbon interstitial migrates along the $[100]$ direction until it recombines with the vacancy. The energy barrier is 1.43~eV, which is in the upper part of the experimental data. 

\begin{figure}
\begin{center}
\includegraphics*[width=8cm]{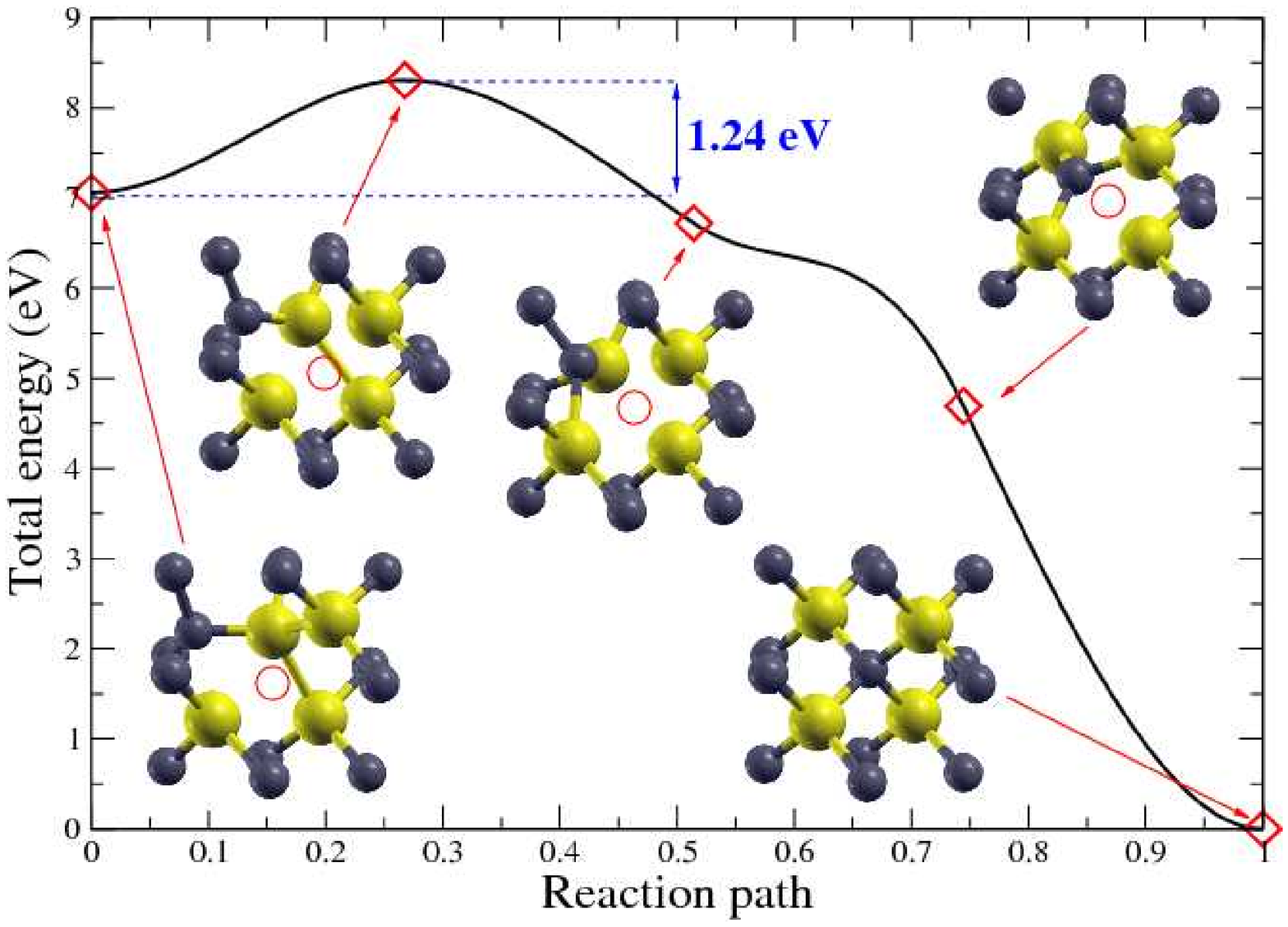}
\caption{Minimum energy path for the recombination of the Frenkel pair $V_C+CSi_{[100]}$ ($d_{FP}=0.5 a_0$). See legend of Figure~\ref{FP1} for further details.} \label{FP2}
\end{center}
\end{figure}

The second studied Frenkel pair is $V_C+CSi_{[100]}$ ($d_{FP}=0.5 a_0$). In that case, the recombination path is not direct, and involves the concerted displacement of the carbon interstitial together with the silicon atom forming the dumbbell (Figure~\ref{FP2}). The calculated energy barrier is 1.24~eV, again in the upper part of the available experimental data. This calculated path and energy barrier are in agreement with a previous calculation, yielding a similar mechanism and an energy barrier of 1.4~eV for an equivalent Frenkel pair configuration~\cite{Boc04PRB}. 

\begin{figure}
\begin{center}
\includegraphics*[width=8cm]{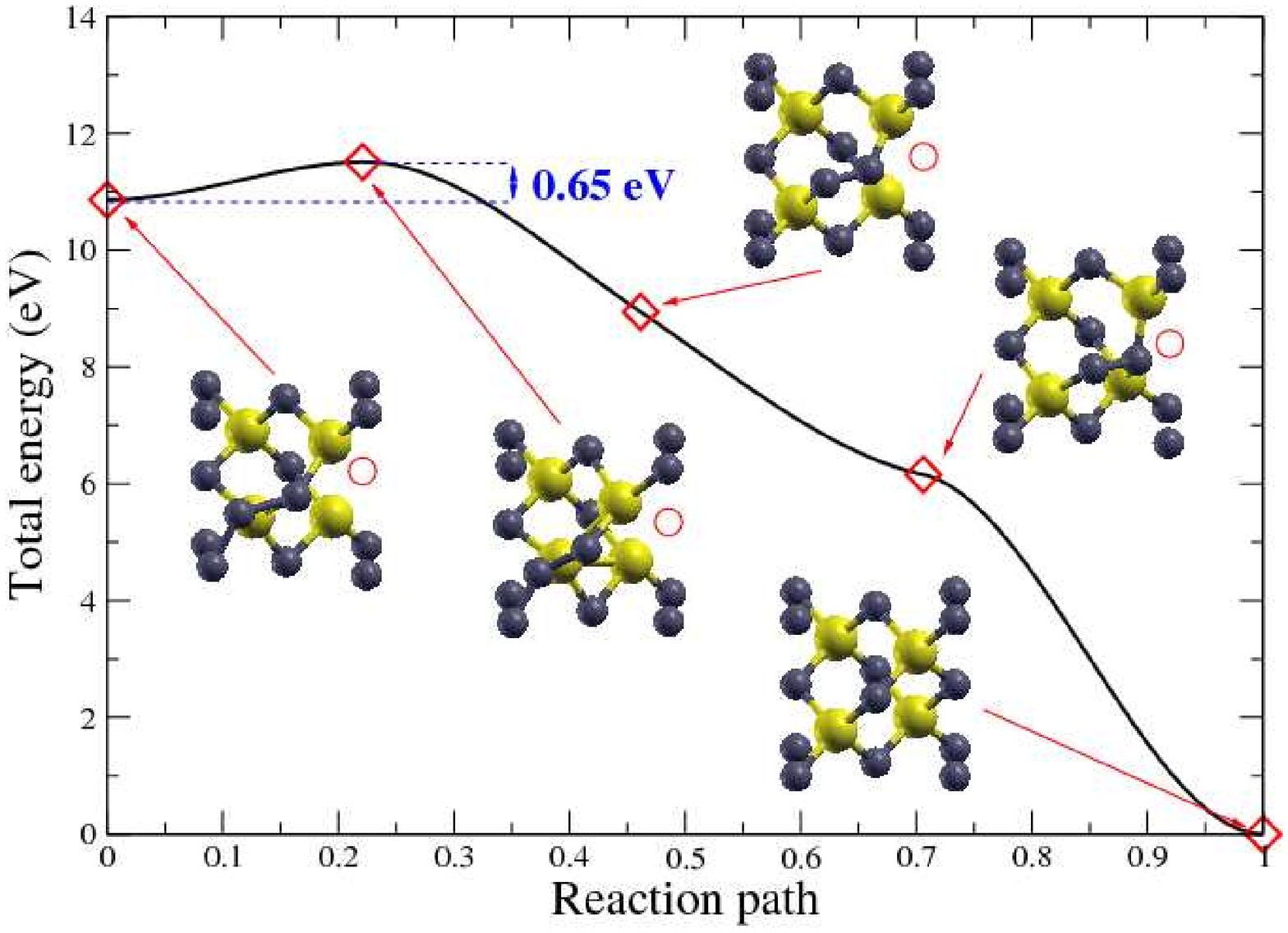}
\caption{Minimum energy path for the recombination of the Frenkel pair $V_C+CSi_{[100]}$ ($d_{FP}=0.95 a_0$). See legend of Figure~\ref{FP1} for further details.} \label{FP3}
\end{center}
\end{figure}

The third configuration, $V_C+CSi_{[100]}$ is close to the previous one, with a larger interstitial-vacancy distance ($d_{FP}=0.95 a_0$). Surprinsingly, the enery barrier is much lower in this case, with 0.65~eV (Figure~\ref{FP3}). This energy is required for the first step of the recombination process, with the transformation of the $CSi$ dumbbell to a $CC$ dumbbell, very close to the vacancy. This $CC$ defect then recombines very easily with the vacancy. Such a result is in agreement with a previous calculation showing a very low recombination energy (0.2~eV) for this last step \cite{Boc04PRB}.  

\begin{figure}
\begin{center}
\includegraphics*[width=8cm]{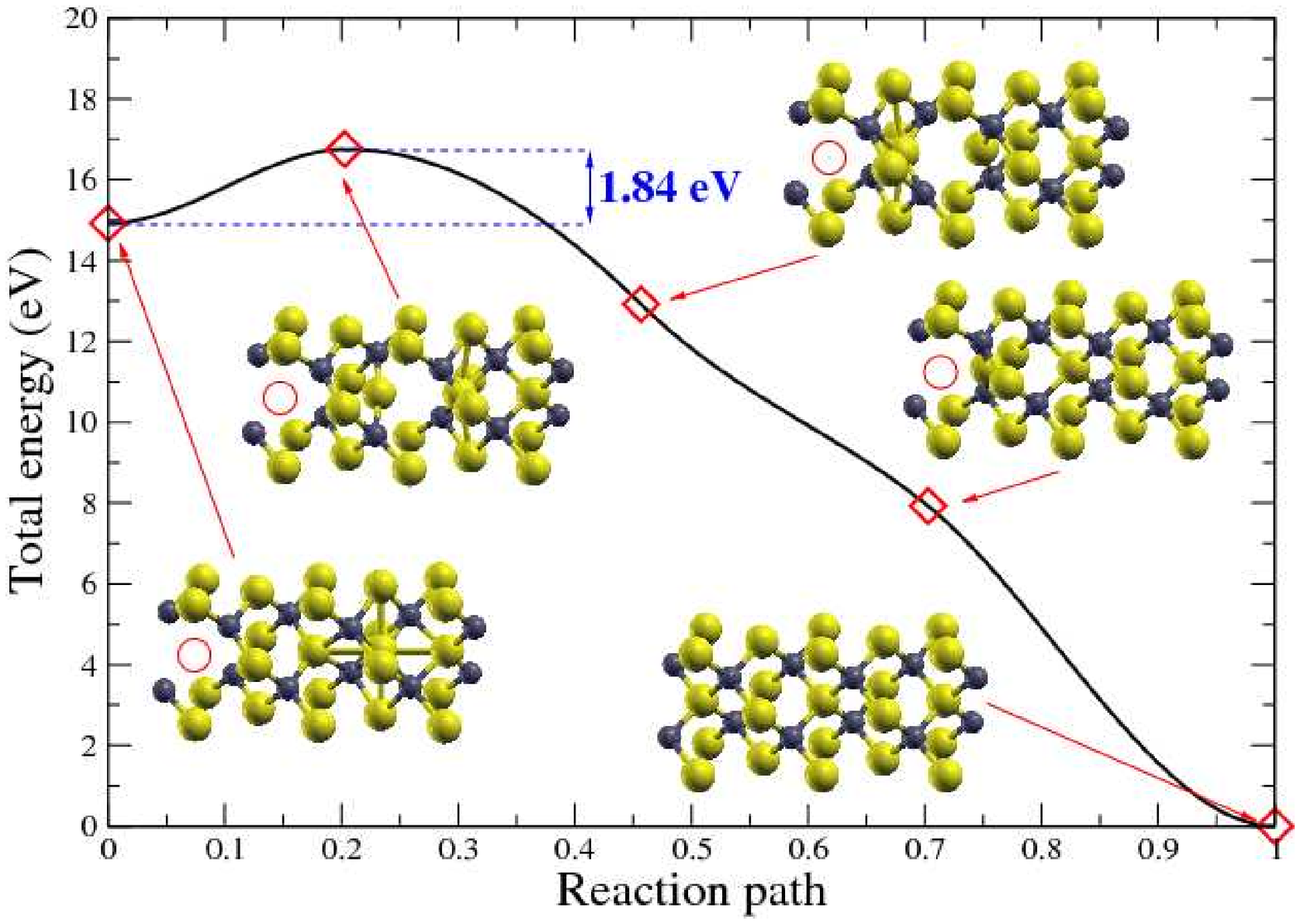}
\caption{Minimum energy path for the recombination of the Frenkel pair $V_{Si}+Si_{TC}$ ($d_{FP}=1.5 a_0$). See legend of Figure~\ref{FP1} for further details.} \label{FP4}
\end{center}
\end{figure}

The Figure~\ref{FP4} shows the recombination of the Frenkel pair $V_{Si}+Si_{TC}$ ($d_{FP}=1.5 a_0$) according to an exchange mechanism. The silicon interstitial replaces another silicon atom, located between the vacancy and the interstitial original position. This silicon atom is filling the vacancy. The whole recombination process is associated with a large energy barrier of 1.84~eV. It is likely that another mechanism, more complex and less expensive in energy, occurred for this configuration. In fact, the initial interstitial-vacancy distance is large, and one may imagine the formation of a $SiSi$ dumbbell, and its subsequent migration toward the vacancy along a longer and easier path. In that case, the barrier for the $SiSi$ migration, computed to be 1.4~eV~\cite{Boc03PRB}, may be the upper energy limit for the recombination. We have investigated such a possibility, by computing the energy barrier for converting the $Si_{TC}$ shown in the Figure~\ref{FP4} to a $SiSi_{[100]}$ dumbbell. The barrier is 1.38~eV, in very close agreement to the calculated energy for the $SiSi$ migration~\cite{Boc03PRB}. After few steps, the interstitial could recombine with the vacancy along the $[110]$ direction, what requires only 0.2~eV~\cite{Boc04PRB}. 

\begin{figure}
\begin{center}
\includegraphics*[width=8cm]{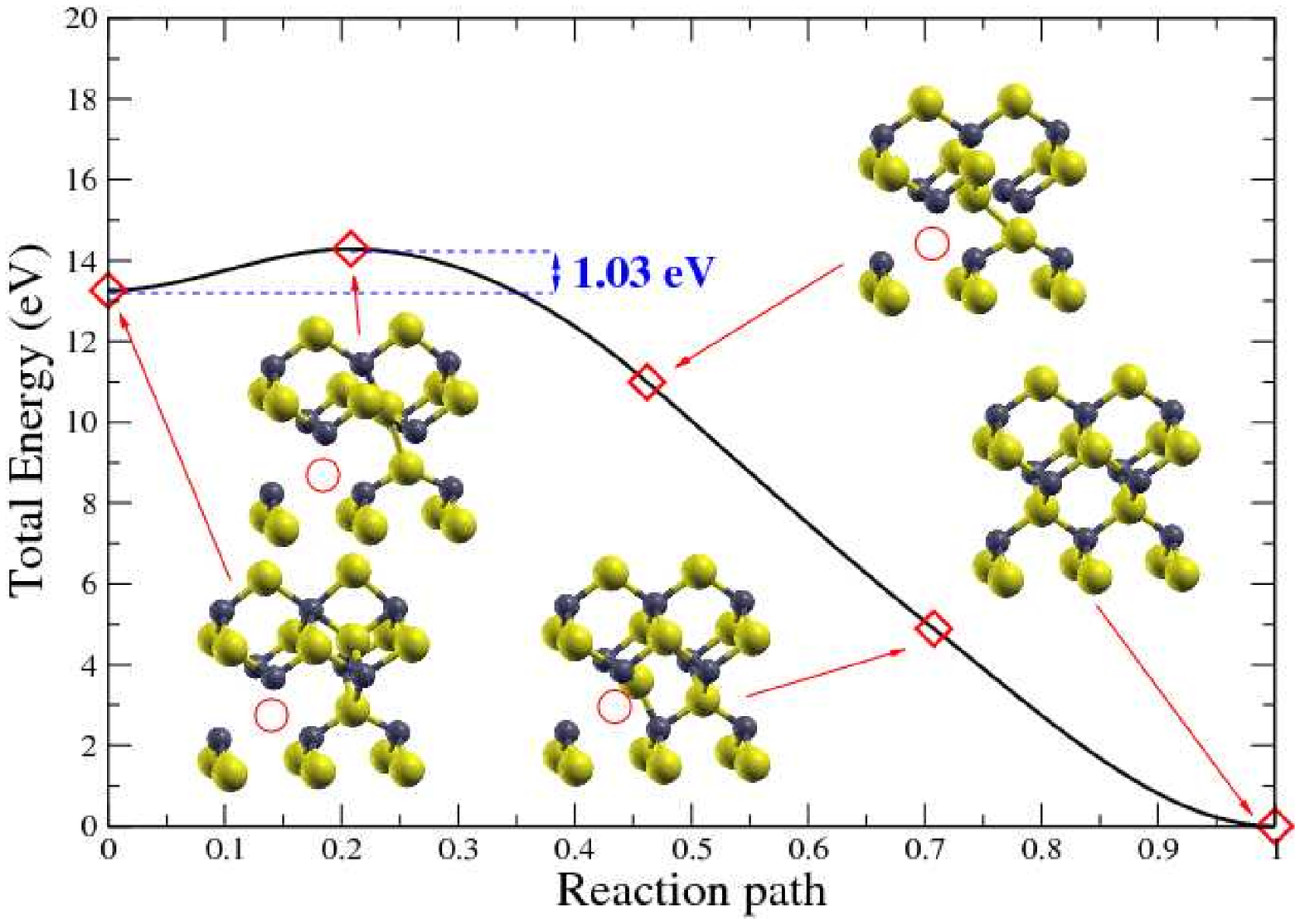}
\caption{Minimum energy path for the recombination of the Frenkel pair $V_{Si}+Si_{TC}$ ($d_{FP}=0.9 a_0$). See legend of Figure~\ref{FP1} for further details.} \label{FP5}
\end{center}
\end{figure}

The last Frenkel pair configuration, $V_{Si}+Si_{TC}$ ($d_{FP}=0.9 a_0$), includes the same interstitial than the previous one, but with a different local geometry (Figure~\ref{FP5}). Here, a trivial recombination mechanism seems the best choice, with a simple straight migration of the silicon interstitial through an hexagonal transition state. The associated activation energy amounts to 1.03 eV.  We also investigated another possible recombination mechanism, in which the interstitial and a silicon neighbour are exchanged. However, a very large activation energy of 2.37~eV makes this event very unlikely. 

\section{Average lifetime of a Frenkel pair}

\begin{figure}
\begin{center}
\includegraphics*[width=8cm]{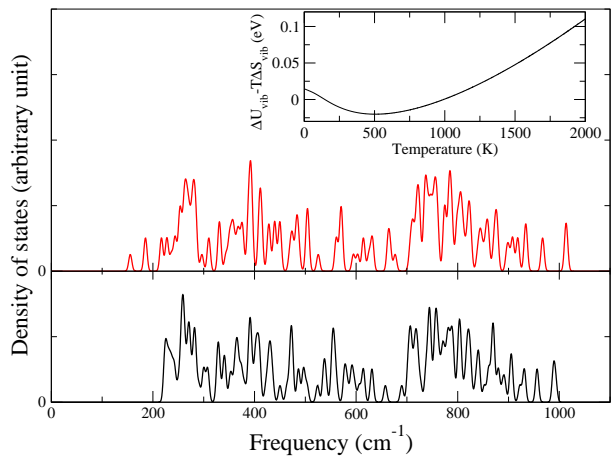}
\caption{Computed phonons frequencies for the Frenkel pair $V_{Si}+Si_{TC}$ (bottom) and the transition state (top) configurations, both shown in the Figure~\ref{FP5}. The inset shows the vibrational contributions to the free energy barrier as a function of temperature.} \label{vib}
\end{center}
\end{figure}

In the framework of harmonic transition state theory, the average lifetime of a Frenkel pair is given by $\tau=(1/\nu_0)e^{-\Delta G/kT}$. $\nu_0$ is the effective attempt frequency, while $\Delta G$ is the free energy difference between initial and transition states. Assuming that the most important contributions are the defect formation energy and vibrational terms, we obtain $\Delta G=E_f+\Delta U_{vib}-T\Delta S_{vib}$. Using the model of harmonic oscillators, vibrational quantities are easily calculated from the following expressions

\begin{equation}
U_{vib}=\sum_{i=1}^{3N}\left[\frac{\hbar\omega_i}{\exp(\hbar\omega_i/kT)-1}+\frac{1}{2}\hbar\omega_i\right]
\end{equation}
\begin{equation}
S_{vib}=k\sum_{i=1}^{3N}\left[\frac{\hbar\omega_i}{kT}\left(\exp(\hbar\omega_i/kT)-1\right)^{-1}-\ln\left(1-\exp(\hbar\omega_i/kT)\right)\right]
\end{equation}

We have determined these quantities in the case of the recombination of the Frenkel pair $V_{Si}+Si_{TC}$ ($d_{FP}=0.9 a_0$). The phonons frequencies $\omega_i$ have been calculated in the frozen phonon approximation for both the initial and transition states obtained in a 64 atoms cell. The figure~\ref{vib} shows the computed frequencies, as well as the vibrational contributions to the free energy barrier as a function of the temperature. Our results indicate that vibrational contributions may be neglected for temperatures usually considered in experiments. When the temperature is as high as 2000~K, the activation energy is increased by 0.1~eV. However, the harmonic approximation is maybe not valid in this regime. 

Using the calculated phonon frequencies for the Frenkel pair (FP) and the transition state (TS) configurations, the effective attempt frequency is obtained from
\begin{equation}
\nu_0=\frac{\prod_{i=1}^{3N}\omega_i^{FP}}{\prod_{i=1}^{3N-1}\omega_i^{TS}}
\end{equation}
The computed value is $\nu_0=298.6$~cm$^{-1}=8.95\times10^{12}$~Hz. Using this value and the free energy barrier for transition, the average lifetime of the Frenkel pair $V_{Si}+Si_{TC}$ is a few hours at 300~K, but drops to few $\mu$s at 600~K. This result is completely in agreement with experiments, showing defects annealing with activation energy around 1.3~eV for temperatures higher than 450~K~\cite{Web01NIMB}.  

\section{Discussion}

The different recombination mechanisms investigated in this work yield activation energies ranging from 0.65~eV to 1.84~eV. The lowest energy is obtained for the annealing of a C interstitial, while the largest corresponds to a Si interstitial annealing. This result is coherent with the greater mobility of C interstitials in silicon carbide. 
We found that there is a large diversity of possible recombination mechanisms. While simple interstitial migration to the vacancy is favoured in some case, lowest recombination energies are often associated with more complex mechanisms, involving exchange of atoms or concerted displacements. Therefore, simple transition state determination methods using constrained paths should be avoided for investigating recombination. 

In this work, we have considered cubic SiC, while most of available experimental data concern common hexagonal polytypes such as 4H and 6H. However, the influence of the polytypism on the recombination mechanism remains unclear. In a recent study, Posselt \textit{et al.} have compared simple point defects in 4H and 3C structures~\cite{Pos04JPCM}. They have shown that, despite additional stable configurations in hexagonal crystals, structure and formation energy are similar for a large majority of defects, due to an equivalent local atomic environment. However, one may expect differences in the recombination paths, especially for Frenkel pairs with large interstitial-vacancy distances. In fact, beyond the first neighbours, hexagonal and cubic structures differ. Nevertheless, it is likely that similar recombination mechanisms will occur in both structures, because of the equivalent local atomic environment, and that activation energies will be close. 

Experimentally, several temperature stages and activation energies for annealing defects have been determined. The lowest measured energy is $0.3\pm0.15$~eV. It may be related to the migration of $CC$ dumbbells, requiring an energy of 0.5~eV~\cite{Boc03PRB}, and subsequent recombination along low energy paths. More probably, it may be related to Frenkel pairs recombination with short separation distance. In fact, Bocsktedte \textit{et al.} have calculated the recombination of two Frenkel pairs with energies equal to 0.2 and 0.4~eV~\cite{Boc04PRB}. These specific configurations were not considered in this work since we have restricted our calculations to Frenkel pairs obtained from displacement energies determinations. Only low-index crystallographic directions are used in such calculations, to obtain the extreme values of the displacement energies range. However, it is possible that other directions allow the formation of Frenkel pairs with mid-range displacement energies but with very low recombination energy barriers. These Frenkel pairs will recombine during the first experimental stage. The second and third stages are associated with energies of $1.3\pm0.25$~eV and $1.5\pm0.3$~eV, respectively. Our calculated barriers fit into these ranges. These stages are then related to annealing of short distance Frenkel pairs, or to the long-range migration of interstitials~\cite{Boc04PRB}. Finally, an activation energy of $2.2\pm0.3$~eV has been reported~\cite{Ito89JAP}, larger than our computed values. It is likely that this stage is associated with the annealing of more complex defects.

\ack
This work was funded by the joint research program "ISMIR" between CEA and CNRS.

\section*{References}
\bibliography{biblio.bib}

\begin{thebibliography}{10}

\bibitem{Cho97MRSB}
W.~J. Choyke and G.~Pensl.
\newblock Physical properties of sic.
\newblock {\em Mater. Res. Soc. Bull.}, 22(3):25, 1997.

\bibitem{Wen98NIMB}
E.~Wendler, A.~Heft, and W.~Wesch.
\newblock Ion-beam induced damage and annealing behaviour in sic.
\newblock {\em Nucl. Instrum. Methods Phys. Res.\ B}, 141:105, 1998.

\bibitem{Sko96NIMPR}
W.~Skorupa, V.~Heera, Y.~Pacaud, and H.~Weishart.
\newblock Ion beam processing of single crystalline silicon carbide.
\newblock {\em Nucl. Instrum. Methods Phys. Res.\ B}, 120:114, 1996.

\bibitem{Hal02NIMPR}
A.~Hall{\'e}n, M.~S. Janson, A.~Yu. Kuznetsov, D.~{\AA}berg, M.~K. Linnarsson,
  B.~G. Svensson, P.~O. Persson, F.~H.~C. Carlsson, L.~Storasta, J.~P. Bergman,
  S.~G. Sridhara, and Y.~Zhang.
\newblock Ion implantation of silicon carbide.
\newblock {\em Nucl. Instrum. Methods Phys. Res.\ B}, 186:186, 2002.

\bibitem{Zha02JAP}
Y.~Zhang, W.~J. Weber, W.~Jiang, A.~Hall{\'e}n, and G.~Possnert.
\newblock Damage evolution and recovery on both si and c sublattices in
  al-implanted 4h-sic studied by rutherford backscattering spectroscopy and
  nuclear reaction analysis.
\newblock {\em J. Appl. Phys.}, 91(10):6388, 2002.

\bibitem{Web00NIMB}
W.~J. Weber, W.~Jiang, and S.~Thevuthasan.
\newblock Defect annealing kinetics in irradiated 6h-sic.
\newblock {\em Nucl. Instrum. Methods Phys. Res.\ B}, 166-167:410, 2000.

\bibitem{Web01NIMB}
W.~J. Weber, W.~Jiang, and S.~Thevuthasan.
\newblock Accumulation, dynamic annealing and thermal recovery of
  ion-beam-induced disorder in silicon carbide.
\newblock {\em Nucl. Instrum. Methods Phys. Res.\ B}, 175-177:26, 2001.

\bibitem{Kuz03JAP}
A.~Yu. Kuznetsov, J.~Wong-Leung, A.~Hall{\'e}n, C.~Jagadish, and B.~G.
  Svensson.
\newblock Dynamic annealing in ion implanted sic: Flux versus temperature
  dependence.
\newblock {\em J. Appl. Phys.}, 94(11):7112, 2003.

\bibitem{Ito89JAP}
H.~Itoh, N.~Hayakawa, I.~Nashiyama, and E.~Sakuma.
\newblock Electron spin resonance in electron-irradiated 3c-sic.
\newblock {\em J. Appl. Phys.}, 66:4529, 1989.

\bibitem{Per00JNM}
J.~M. Perlado, L.~Malerba, A.~S{\'a}nchez-Rubio, and T.~D{\'\i}az de~la Rubia.
\newblock Analysis of displacement cascades and threshold displacement energies
  in $\beta$-sic.
\newblock {\em J. Nucl. Mater.}, 276:235, 2000.

\bibitem{Dev01JAP}
R.~Devanathan, W.~J. Weber, and F.~Gao.
\newblock Atomic scale simulation of defect production in irradiated 3c-sic.
\newblock {\em J. Appl. Phys.}, 90(5):2303, 2001.

\bibitem{Boc03PRB}
M.~Bockstedte, A.~Mattausch, and O.~Pankratov.
\newblock Ab initio study of the migration of intrinsic defects in 3c-sic.
\newblock {\em Phys. Rev. B}, 68:205201, 2003.

\bibitem{RauPHD}
E. Rauls, PhD, Universit{\"a}t Paderborn, Germany (2003).

\bibitem{Sal04JNM}
M.~Salvador, J.~M. Perlado, A.~Mattoni, F.~Bernardini, and L.~Colombo.
\newblock Defect energetics of $\beta$-sic using a new tight-binding molecular
  dynamics model.
\newblock {\em J. Nucl. Mater.}, 329-333:1219, 2004.

\bibitem{Gao03JAP}
F.~Gao and W.~J. Weber.
\newblock Recovery of close frenkel pairs produced by low energy recoils in
  sic.
\newblock {\em J. Appl. Phys.}, 94(7):4348, 2003.

\bibitem{Mal00JNM}
L.~Malerba, J.~M. Perlado, A.~S{\'a}nchez-Rubio, I.~Pastor, L.~Colombo, and
  T.~Diaz de~la Rubia.
\newblock Molecular dynamics simulation of defect production in irradiated
  $\beta$-sic.
\newblock {\em J. Nucl. Mater.}, 283-287:794, 2000.

\bibitem{Mal02PRB}
L.~Malerba and J.~M. Perlado.
\newblock Basic mechanisms of atomic displacement production in cubic silicon
  carbide: A molecular dynamics study.
\newblock {\em Phys. Rev. B}, 65:45202, 2002.

\bibitem{Boc04PRB}
M.~Bockstedte, A.~Mattausch, and O.~Pankratov.
\newblock Ab initio study of the annealing of vacancies and interstitials in
  cubic sic: Vacancy-interstitial recombination and aggregation of carbon
  interstitials.
\newblock {\em Phys. Rev. B}, 69:235202, 2004.

\bibitem{Jon98WS}
H.~J{\'o}nsson, G.~Mills, and K.~W. Jacobsen.
\newblock Nudged elastic band method for finding minimum energy paths of
  transitions.
\newblock In B.~J. Berne, G.~Ciccotti, and D.~F. Coker, editors, {\em Classical
  and Quantum Dynamics in Condensed Phase Simulations}, chapter~16, page 385.
  World Scientific, 1998.

\bibitem{Hen00JCP}
G.~Henkelman, B.~P. Uberuaga, and H.~J{\'o}nsson.
\newblock A climbing image nudged elastic band method for finding saddle points
  and minimum energy paths.
\newblock {\em J. Chem. Phys.}, 113(22):9901, 2000.

\bibitem{Luc05PRB}
G.~Lucas and L.~Pizzagalli.
\newblock Ab initio molecular dynamics calculations of threshold displacement
  energies in silicon carbide.
\newblock {\em Phys. Rev. B}, 72:161202R, 2005.

\bibitem{Quantum-Espresso}
Quantum-Espresso package, http://www.quantum-espresso.org.

\bibitem{Hoh64PR}
P.~Hohenberg and W.~Kohn.
\newblock Inhomogeneous electron gas.
\newblock {\em Phys. Rev.}, 136(3B):B864, 1964.

\bibitem{Koh65PR}
W.~Kohn and L.~J. Sham.
\newblock Self-consistent equations including exchange and correlation effects.
\newblock {\em Phys. Rev.}, 140(4A):A1133, 1965.

\bibitem{Per96PRL}
J.~P. Perdew, K.~Burke, and M.~Ernzerhof.
\newblock Generalized gradient approximation made simple.
\newblock {\em Phys. Rev. Lett.}, 77(18):3865, 1996.

\bibitem{Mon76PRB}
H.~J. Monkhorst and J.~D. Pack.
\newblock Special points for brillouin-zone integrations.
\newblock {\em Phys. Rev. B}, 13(12):5188, 1976.

\bibitem{Luc06NIMB}
G.~Lucas and L.~Pizzagalli, {\em Nucl. Instrum. Methods Phys. Res.\ B}, in
  print (2006).

\bibitem{Zha91PRL}
S.~B. Zhang and J.~E. Northrup.
\newblock Chemical potential dependence of defect formation energies in gaas:
  Application to ga self-diffusion.
\newblock {\em Phys. Rev. Lett.}, 67(17):2339, 1991.

\bibitem{Pos04JPCM}
M.~Posselt, F.~Gao, W.~J. Weber, and V.~Belko.
\newblock A comparative study of the structure and energetics of elementary
  defects in 3c- and 4h-sic.
\newblock {\em J. Phys.: Condens. Matter}, 16:1307, 2004.

\end{thebibliography}
\bibliographystyle{unsrt}

\end{document}